\documentclass[aps,prb,twocolumn,showpacs,superscriptaddress]{revtex4-2}

\usepackage{graphicx}
\usepackage{amsmath}
\usepackage{multirow}
\usepackage{tabularx}
\usepackage{color}
\usepackage[colorlinks,linkcolor=blue,urlcolor=blue,anchorcolor=blue,citecolor=blue]{hyperref}

\DeclareRobustCommand{\&}{%
   \ifdim\fontdimen1\font>0pt
    \textsl{\symbol{`\&}}%
   \else
    \symbol{`\&}%
   \fi
}

\begin{document}

\title{Tuning magnetic anisotropy in Fe$_{5}$GeTe$_{2}$ monolayer through doping and strain}
	
\author{Xunwu Hu}

\affiliation{Center for Neutron Science and Technology, Guangdong Provincial Key Laboratory of Magnetoelectric Physics and Devices, State Key Laboratory of Optoelectronic Materials and Technologies, School of Physics, Sun Yat-Sen University, Guangzhou, 510275, China}
\affiliation{Department of Physics, College of Physics \& Optoelectronic Engineering, Jinan University, Guangzhou, 510632, China}
\author{Dao-Xin Yao}
\email{yaodaox@mail.sysu.edu.cn}
\affiliation{Center for Neutron Science and Technology, Guangdong Provincial Key Laboratory of Magnetoelectric Physics and Devices, State Key Laboratory of Optoelectronic Materials and Technologies, School of Physics, Sun Yat-Sen University, Guangzhou, 510275, China}
\author{Kun Cao}
\email{caok7@mail.sysu.edu.cn}
\affiliation{Center for Neutron Science and Technology, Guangdong Provincial Key Laboratory of Magnetoelectric Physics and Devices, State Key Laboratory of Optoelectronic Materials and Technologies, School of Physics, Sun Yat-Sen University, Guangzhou, 510275, China}

\begin{abstract}

Controlling magnetic anisotropy energy (MAE) in two-dimensional (2D) ferromagnetic materials is crucial for designing novel spintronic devices. Using first-principles calculations, we systematically investigate the magnetic properties of monolayer Fe$_5$GeTe$_2$ (F5GT) under two scenarios: (I) Co and Ni doping, and (II) compressive and tensile strains. Our results show that the F5GT monolayer exhibits a weak in-plane MAE, which can be significantly enhanced by Co doping. Additionally, a $1\%$ compressive strain switches the magnetic easy axis from in-plane to out-of-plane, while $4\%$ compressive strain can further enhance the out-of-plane MAE. Spin-orbit coupling (SOC) matrix analysis reveals that the enhancement of in-plane MAE in Co-doped F5GT (Co-F5GT) arises from changes in $ \left\langle {\it p}_{x} \left\vert L_z\right\vert {\it p}_{y} \right\rangle  $ of Te and  $ \left\langle {\it d}_{xy} \left\vert L_z\right\vert {\it d}_{x^2+y^2} \right\rangle  $ of Fe(2) and Fe(3). The effect of compressive strain is primarily attributed to a substantial increase in the positive contribution from $ \left\langle {\it d}_{xy} \left\vert L_z\right\vert {\it d}_{x^2+y^2} \right\rangle  $ of Fe(1). 
 
\end{abstract}
\maketitle

\section{Introduction}

Since the discovery of 2D ferromagnets Cr$_{2}$Ge$_{2}$Te$_{6}$~\cite{2017gong} and CrI$_{3}$~\cite{2017huan}, 2D van der Waals (vdWs) magnetic materials have been attracting great amount of attention~\cite{JL2019,kong2019,2006ejicDeiseroth,2013jpsjChen,2016prbMay,2019prlJohansen,Fe3,2020sciadv,2021prbMondal}, owing to their prospective applications in next-generation ultra-thin magnetic memory and spintronic devices. However, practical applications of 2D magnetic materials are usually prevented by their low ordering temperatures $T_c$. To show a few examples, both Cr$_{2}$Ge$_{2}$Te$_{6}$~\cite{2017gong} and CrI$_{3}$~\cite{2017huan} have $T_c \approx 60$~K. Although ferromagnetic metals Fe$_{3}$GeTe$_{2}$~\cite{2006ejicDeiseroth,2013jpsjChen,2016prbMay} and Fe$_{4}$GeTe$_{2}$~\cite{2020sciadv} have higher  $T_c \approx$~230 K  and $T_c \approx$ 270 K, respectively. The $T_c$'s are still below room temperature, especially in their monolayer or few layer form. To enable high-quality and wide-range applications, it is desired that the magnets remain functional beyond room temperature. Bulk F5GT exhibits metallic ferromagnetism near room temperature~\cite{2019prm,2019acsnano,2020prbzhang,20212dm,Chen2022}, with an increased $T_c$ of 328 - 350 K in Co-F5GT \cite{2022sciav,chen2023above,2020prm,2020apl,2022prmz}.  Notably, a significant enhancement of the $T_c$ to 478 K has been reported with 36$\%$ Ni doping~\cite{2022prl}. Apart from its bulk form, ferromagnetism near or above room temperature has also been observed in thin flakes of F5GT and Co-F5GT ~\cite{20212dm,Chen2022,2022sciav,chen2023above}. Our previous theoretical work predicts ferromagnetism with $T_c$ = 430~K in 20$\%$ Ni-doped F5GT~(Ni-F5GT) monolayer~\cite{huprb}. Therefore, the F5GT system is an ideal candidate for practical applications in spintronic devices.

On the other hand, a sizeable MAE is crucial not only for stabilizing 2D magnetic ordering but also for the developing advanced spintronic devices, such as spin valve~\cite{zhaospinvalve} and magnetic tunnel junction~\cite{wupra2023}. However, both experimental and theoretical studies suggest a relatively weak MAE in F5GT systems~\cite{2019acsnano,2019prm,20212dm,2021zhaoyu,2021prbyao,Chen2022,2020prbzhang,2022jpcl}. Fortunately, doping and strain engineering have proven effective in modulating the MAE of 2D magnetic materials and their vdWs heterostructures, as demonstrated in various research~\cite{2020yinprb,2017yang,2019dongpra,2018webster,2022tang}. Thus, manipulating MAE through doping and strain presents a promising avenue for enhancing the magnetic properties of F5GT and advancing its applicability in spintronic applications.

In this paper, using first-principles calculations, we systematically investigate the magnetic properties of monolayer F5GT under two scenarios: (I) Co and Ni doping, and (II) in-plane strains with $-5\% \le \varepsilon \le +5\%$. Our results show that F5GT monolayer exhibits a weak in-plane magnetic anisotropy, which can be significantly enhanced by Co doping, resulting in a threefold increase from -0.58 meV/f.u. to -1.51 meV/f.u. Notably, a sign change in MAE occurs at $ \varepsilon = -1\%$, indicating a transformation from in-plane to out-of-plane  magnetization, while $4\%$ compressive strain further enhances the out-of-plane MAE. SOC matrix analysis reveals that the enhancement of in-plane MAE in Co-F5GT is attributed to changes in $ \left\langle {\it p}_{x} \left\vert L_z\right\vert {\it p}_{y} \right\rangle  $ of Te and  $ \left\langle {\it d}_{xy} \left\vert L_z\right\vert {\it d}_{x^2+y^2} \right\rangle  $  of Fe(2) and Fe(3). The effect of compressive strain is primarily due to a significant increase in the positive contribution from  $ \left\langle {\it d}_{xy} \left\vert L_z\right\vert {\it d}_{x^2+y^2} \right\rangle  $ of Fe(1). These findings highlight the tunability of magnetic properties in F5GT, offering insights for future spintronic applications.

\begin{figure}[t]
	\includegraphics[scale=0.4]{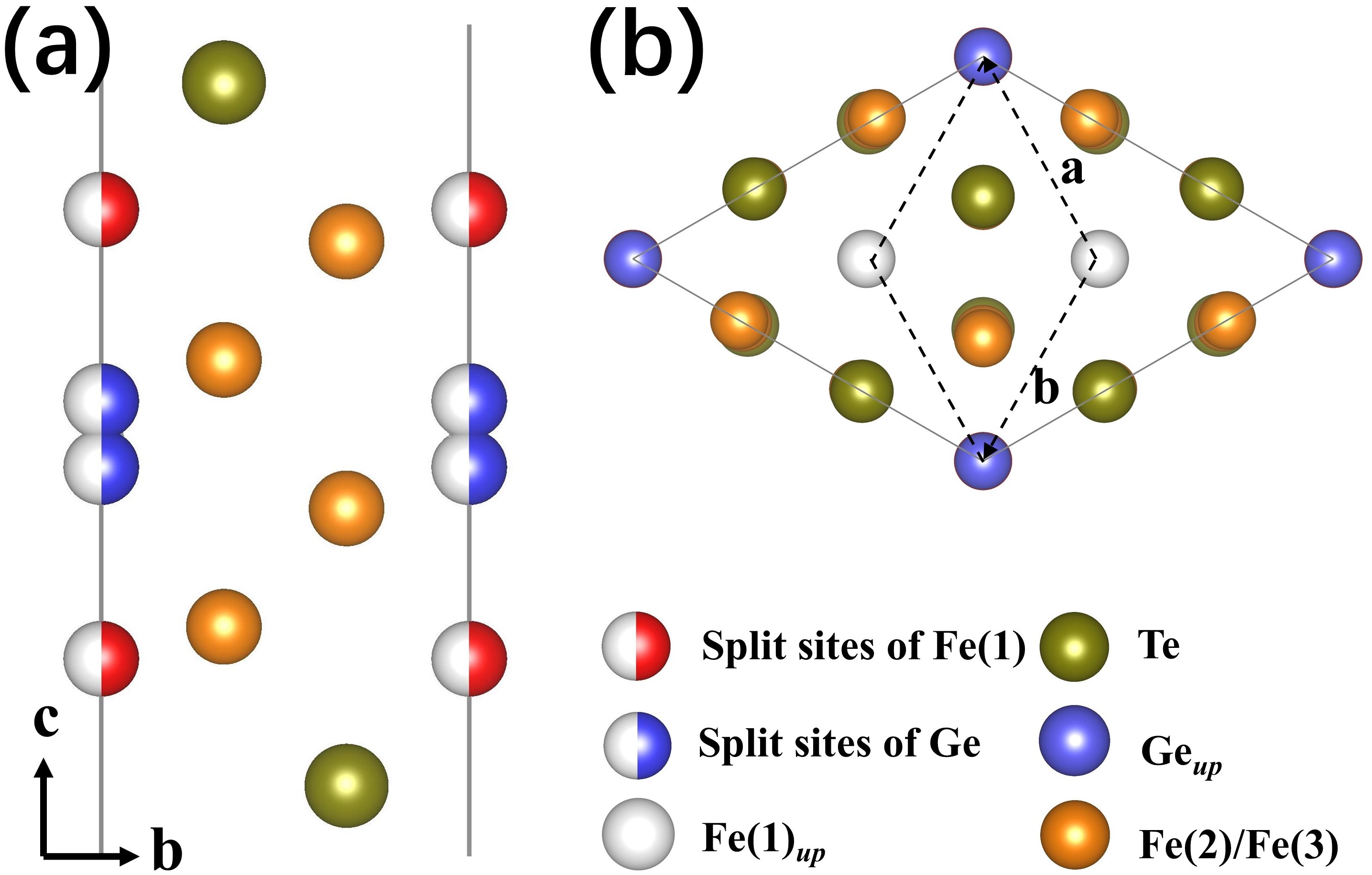}
	\caption{(a) Side view of the average crystal structure of the F5GT monolayer. Half-white, half-red spheres indicate the split sites of Fe(1), while orange and dark green spheres represent Fe(2)/(3) and Te atoms, respectively. (b) Top view of the F5GT monolayer with a $\sqrt{3}\times\sqrt{3}$ superstructure. White and blue spheres correspond to Fe(1)$_{up}$ and Ge$_{up}$ atoms, respectively, with Fe(1)$_{dn}$ positioned directly beneath the Ge$_{up}$ atom. The primitive cell is outlined by a black dashed line.}
	\label{fig1}
\end{figure}

\section{Computational details}
Our first-principles calculations are performed using density functional theory as implemented in the Vienna {\it ab initio} simulation package (VASP)~\cite{93prb,96prb}. The generalized gradient approximation (GGA) of Perdew-Burke-Ernzerhof (PBE)~\cite{96prl} form is used for the exchange-correlation functional. The projector augmented-wave (PAW) method~\cite{94prb}, with a 400 eV plane-wave cutoff, is employed. The in-plane lattice parameters and atomic positions are fully optimized until forces on each atom are less than $10^{-2} $ eV/$\mathring{A}$. During the electronic self-consistent steps, the energy convergence criterion is set to $10^{-6} $ eV. The out-of-plane lattice parameter is fixed at 22.6 $\mathring{A}$, with vacuum spacing exceeding 16 $\mathring{A}$, to avoid interactions between adjacent images.

The MAE is defined as $E_{\rm MAE }$ = $E_{\rm[100]}$ - $E_{\rm[001]}$,
where $E_{\rm[100]}$ and $E_{\rm[001]}$ represent the calculated energies with in-plane [100] and out-of-plane [001] magnetization directions, respectively. A positive $E_{\rm MAE}$ implies an out-of-plane
magnetic easy axis, while a negative value indicates an in-plane easy axis. It is worth noting that the total MAE contains the magnetocrystalline anisotropy (MCA) energy ($E_{\rm MCA}$) and the magnetic shape anisotropy (MSA) energy ($E_{\rm MSA}$), expressed as $E_{\rm MAE}$ = $E_{\rm MCA}$ + $E_{\rm MSA}$. 

The $E_{\rm MCA}$, arising from SOC, is evaluated using $E_{\rm MCA }$ = $E_{\rm[100]}^{\rm SOC}$ - $E_{\rm[001]}^{\rm SOC}$. Based on the second-order perturbation theory~\cite{1993prbWang}, the $E_{\rm MCA }$~can be approximated as,
\begin{eqnarray}\label{eq:eq1}
	E_{\rm MCA}=\xi^2\sum_{o,u,\sigma}\frac{{\left\vert \left\langle o,\sigma\left\vert L_z\right\vert u,\sigma\right\rangle  \right\vert} ^2-{\left\vert \left\langle o,\sigma\left\vert L_x\right\vert u,\sigma\right\rangle  \right\vert} ^2}{E_{u,\sigma} - E_{o,\sigma}}\notag\\   
	- \xi^2\sum_{o,u,\sigma \neq \sigma'}\frac{{\left\vert \left\langle o,\sigma\left\vert L_z\right\vert u,\sigma'\right\rangle  \right\vert} ^2-{\left\vert \left\langle o,\sigma\left\vert L_x\right\vert u,\sigma'\right\rangle  \right\vert} ^2}{E_{u,\sigma'} - E_{o,\sigma}}
\end{eqnarray}
where $\xi$ is the radial part of SOC, $L_z$ and $L_x$ are the angular momentum operators for out-of-plane and in-plane directions, respectively. ${\left\vert \left\langle o,\sigma\left\vert L_z\right\vert u,\sigma\right\rangle  \right\vert} ^2-{\left\vert \left\langle o,\sigma\left\vert L_x\right\vert u,\sigma\right\rangle  \right\vert} ^2$~and~${\left\vert \left\langle o,\sigma\left\vert L_z\right\vert u,\sigma'\right\rangle  \right\vert} ^2-{\left\vert \left\langle o,\sigma\left\vert L_x\right\vert u,\sigma'\right\rangle  \right\vert} ^2$~represent the matrix element differences between out-of-plane and in-plane magnetization directions. $\left\vert o, \sigma \right\rangle$ and $\left\vert u, \sigma \right\rangle$ indicate occupied and unoccupied states with spin $\sigma$, respectively, with corresponding energies $E_{o,\sigma}$ and $E_{u,\sigma}$. Given the sensitivity of $E_{\rm MCA}$ to the choice of {\it k}-point grid, we carefully check the convergence of $E_{\rm MCA}$ in the F5GT monolayer. All results reported in this work are calculated using a dense $k$-mesh of ~14 $\times $ 14 $\times $ 1, ensuring convergence of $E_{\rm MCA }$ within 0.005 meV/f.u.

The $E_{\rm MSA}$, resulting from long-range magnetic dipole-dipole interactions, is calculated using the formula $E_{\rm MSA }$ = $E_{\rm[100]}^{\rm Dipole}$ - $E_{\rm[001]}^{\rm Dipole}$, where $E^{\rm Dipole }$ can be expressed as
\begin{eqnarray}
	E^{\rm Dipole}= \frac{1}{2} \frac{\mu_{0}}{4\pi} \sum_{i \neq j} \left [ \frac{\textbf{M}_{i} \cdot \textbf{M}_{j} } {r_{ij}^{3}} - \frac{3}{r_{ij}^{5}} (\textbf{M}_{i} \cdot \textbf{r}_{ij}) (\textbf{M}_{j} \cdot \textbf{r}_{ij})  \right],
\end{eqnarray}
where $\textbf{M}_{i}$ and $\textbf{M}_{j}$ are the local magnetic moments, and $\textbf{r}_{ij}$ donates the vector pointing from site $i$ to site $j$. $\mu_{0}$ represents the vacuum permeability. A cutoff of $r_{ij} = 1200~\mathring{A}$ is adopted to ensure the convergence of $E_{\rm MSA }$.

\section{Results and discussion}

\subsection{Structural properties}
The bulk F5GT exhibits a trigonal average crystal structure with the space group R$\bar{3}$m~\cite{2019acsnano}, comprising three identical monolayer blocks within a unit cell. Each monolayer contains a Fe$_5$Ge sublayer sandwiched between two Te planes. Within the outermost plane of the Fe$_5$Ge sublayer, Fe(1) atoms can occupy one of two possible split sites, labled as Fe(1)$_{up}$ and Fe(1)$_{dn}$, positioned either above or below the Ge atom, as illustrated in Fig. \ref{fig1}(a). This unique structural feature distinguishes F5GT from other Fe$_n$GeTe$_2$ compounds, such as Fe$_3$GeTe$_2$ and Fe$_4$GeTe$_2$.

Given the complexity of managing structural disorder induced by the split sites, most theoretical studies have focused on ordered lattice structures in which Fe(1) atoms exclusively occupy either the up~{\it uuu} or equivalent down~{\it ddd} positions~\cite{2021prbyao,2021zhaoyu,2022prmz}. However, scanning tunneling microscopy experiments have revealed that the F5GT compound forms ordered $\sqrt{3}\times\sqrt{3}$ superstructures, driven by the ordering within the Fe(1) layer. These configurations include structures such as Fe(1)$_{dn}$Fe(1)$_{up}$Fe(1)$_{up}$ ({\it duu}) or equivalent Fe(1)$_{up}$Fe(1)$_{dn}$Fe(1)$_{dn}$ ({\it udd})~\cite{2021adfm,2021prb}. In addition, our pervious work has demonstrated that {\it duu} or {\it udd} configurations are energetically more favorable than {\it uuu} or {\it ddd} structures~\cite{huprb}. Accordingly, this study considers a $\sqrt{3}\times\sqrt{3}$ {\it duu} superstructure of the F5GT monolayer, featuring one Fe(1) atom positioned downwards and two Fe(1) atoms oriented upwards, as illustrated in Fig. \ref{fig1}(b).

First-principles calculations indicate that, at low concentration, Co and Ni preferentially substitute for Fe(1) atoms in Co-F5GT and Ni-F5GT, repectively~\cite{huprb,2020prm}. However, at higher concentrations, these systems may undergo structural transitions accompanied by more complex occupation patterns. This study focuses on 20$\%$ Co-F5GT and 20$\%$ Ni-F5GT monolayers, where all Fe(1) atoms are replaced by Co or Ni, while the other Fe atoms remain unaffected. For simplicity, these compositions will be referred as Co-F5GT and Ni-F5GT. The fully relaxed in-plane lattice parameters for F5GT, Co-F5GT, and Ni-F5GT monolayers are~7.00~$\mathring{A}$, 6.98~$\mathring{A}$, and 6.98 $\mathring{A}$, respectively, indicating that the substitution of Co and Ni results in a slight contraction of the in-plane lattice.

\subsection{Magnetic moments} 

The strain-dependent magnetic moments (${\textbf m}$) are investigated in this subsection. As listed in Table \ref{tab:table1}, the magnetic moments of Fe(1)$_{up}$, Fe(1)$_{dn}$, Fe(2), and Fe(3) in the unstrained F5GT monolayer are 1.87, 1.90, 2.14, and 2.53 $\mu_B$, respectively. In the Ni-F5GT and Co-F5GT monolayers, the Ni and Co atoms exhibit relatively weak magnetic moments (${\textbf m}_{\rm{Ni}}$~$\sim$~0.26 $\mu_B$ and ${\textbf m}_{\rm{Co_{up}}}$~$\sim$~0.57 $\mu_B$), while the remaining Fe atoms exhibit minimal variation in ${\textbf m}$. Consequently, the total magnetic moments (${\textbf m}_{tot}$) of the unstrianed F5GT monolayer is 11.05 $\mu_B$/f.u., which decreases  to 9.63 $\mu_B$/f.u. for Co-F5GT and 9.60 $\mu_B$/f.u. for Ni-F5GT.
 \begin{figure}[t]
 	\includegraphics[scale=0.33]{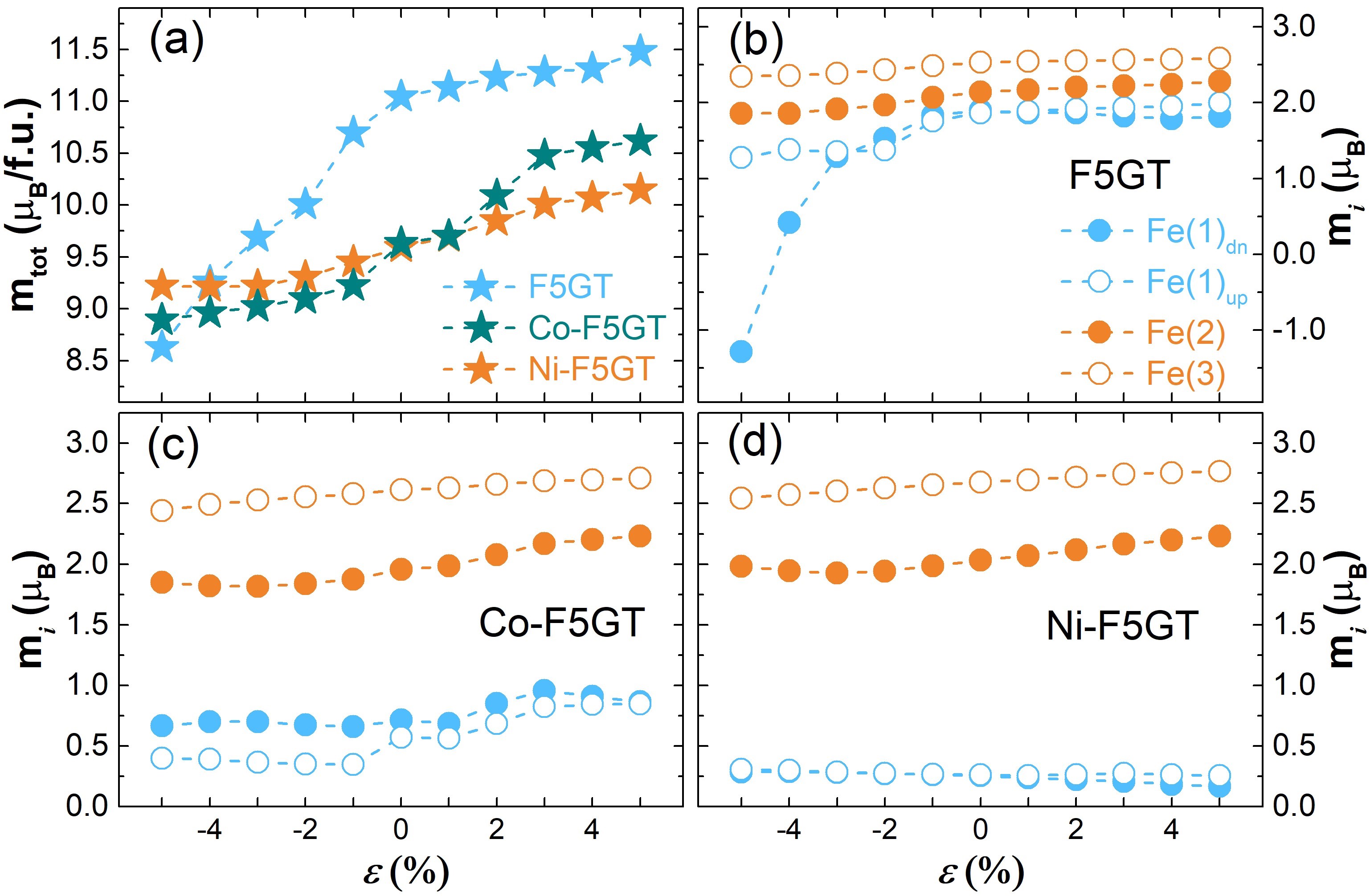}
 	\centering
 	\caption{(a)~The calculated total magnetic moments~${\textbf m}_{tot}$ of F5GT, Co-F5GT, and Ni-F5GT monolayers as a function of strain. Strain-dependent magnetic moments~${\textbf m}_{i}$ at different sites for (b)~F5GT, (c)~Co-F5GT, and (d)~Ni-F5GT monolayers.}
 	\label{fig2}
 \end{figure}

Fig. \ref{fig2} illustrates the effects of strain on the magnetic moments of F5GT, Ni-F5GT, and Co-F5GT monolayers. As shown in Fig. \ref{fig2}(a), ${\textbf m}_{tot}$ decreases under compressive strain and increases under tensile strain. For Ni-F5GT, ${\textbf m}_{tot}$ varies slightly, ranging from 9.2 $\mu_B$/f.u. to 10.0 $\mu_B$/f.u., while for Co-F5GT, it ranges from 9.0 $\mu_B$/f.u. to 10.6 $\mu_B$/f.u. Obviously, the magnetic moments of Ni-F5GT and Co-F5GT monolayers exhibit limited sensitivity to strain, as shown in Figs. \ref{fig2}(c) and \ref{fig2}(d). In contrast, the ${\textbf m}_{tot}$ of F5GT monolayer demonstrates greater sensitivity to external strain, varying from 8.6 $\mu_B$/f.u. to 11.6 $\mu_B$/f.u. Among its components, Fe(1)$_{up}$, Fe(2), and Fe(3) moments show minimal sensitivity to strain. However, Fe(1)$_{dn}$ moment undergoes a sign change, shifting from -1.29 $\mu_B$ to 1.92 $\mu_B$, as depicted in Fig. \ref{fig2}(b). This sign change indicates a transformation in the direction of magnetization, contributing significantly to the increase in ${\textbf m}_{tot}$ for the F5GT monolayer. 
 \begin{table}[t]
	\caption{\label{tab:table1} The calculated magnetic moments ${\textbf m}$$_i$ of unstrained F5GT, Co-F5GT, and Ni-F5GT monolayers. The units are $\mu_B$.}
	\begin{ruledtabular}
		\centering
		\begin{tabular}{cccc}
			\multicolumn{1}{c}{Index}&{F5GT}& {Co-F5GT} & {Ni-F5GT} \\
			\hline
			Fe(1)$_{up}$  &1.87  &0.57 &0.26 \\
			Fe(1)$_{dn}$  &1.90  &0.71 &0.26\\  
			Fe(2)         &2.14  &1.96 &2.03 \\  
			Fe(3)         &2.53  &2.62 &2.68\\   
		\end{tabular}
	\end{ruledtabular}
\end{table}
\subsection{Magnetic anisotropy}
We now investigate the effects of doping and strain on the MAE of the F5GT monolyaer. The calculated $E_{\rm MCA }$, $E_{\rm MSA }$, and $E_{\rm MAE }$ for unstrained F5GT, Ni-F5GT, and Co-F5GT monolayers are listed in Table \ref{tab:table2}. All three systems exhibit negative $E_{\rm MCA }$, indicating a preference for in-plane magnetization alignment. Specifically, the F5GT monolayer shows an extremely weak $E_{\rm MCA }$ of -0.10 meV/f.u., while Ni-F5GT has a slightly larger $E_{\rm MCA }$ of -0.36 meV/f.u. The Co-F5GT monolayer demonstrates a significant enhancement of $E_{\rm MCA }$, reaching -1.15 meV/f.u. This substantial increase in  $E_{\rm MCA }$ for Co-F5GT can be understood in terms of SOC matrix elements, as discussed in section D. Furthermore, all calculated $E_{\rm MSA }$ favor an in-plane easy axis, with $E_{\rm MSA }$ for Ni-F5GT comparable to $E_{\rm MCA }$, suggesting a significant contribution from MSA in the MAE for these 2D magnetic materials~\cite{2019prbXue}. Consequently, the overall $E_{\rm MAE }$ for unstrained F5GT, Ni-F5GT, and Co-F5GT monolayers are -0.58 meV/f.u., -0.71 meV/f.u., and -1.51 meV/f.u., respectively. These results reveal that doping with Ni and Co enhances the MAE of the F5GT monolayer.

\begin{table}[t]
	\caption{\label{tab:table2} The calculated $E_{\rm MCA }$, $E_{\rm MSA }$, and $E_{\rm MAE }$ of unstrained F5GT, Co-F5GT, and Ni-F5GT monolayers. The units are meV/f.u.}
	\begin{ruledtabular}
		\centering
		\begin{tabular}{cccc}
			\multicolumn{1}{c}{Index}&{F5GT}& {Co-F5GT} & {Ni-F5GT} \\
			\hline  
			$E_{\rm MCA }$ &-0.10  &-1.15 &-0.36    \\
			$E_{\rm MSA }$ &-0.48  &-0.36 &-0.35  \\
			$E_{\rm MAE }$ &-0.58  &-1.51 &-0.71  \\    
		\end{tabular}
	\end{ruledtabular}
\end{table}
 
\begin{figure}[t]
	\includegraphics[scale=0.47]{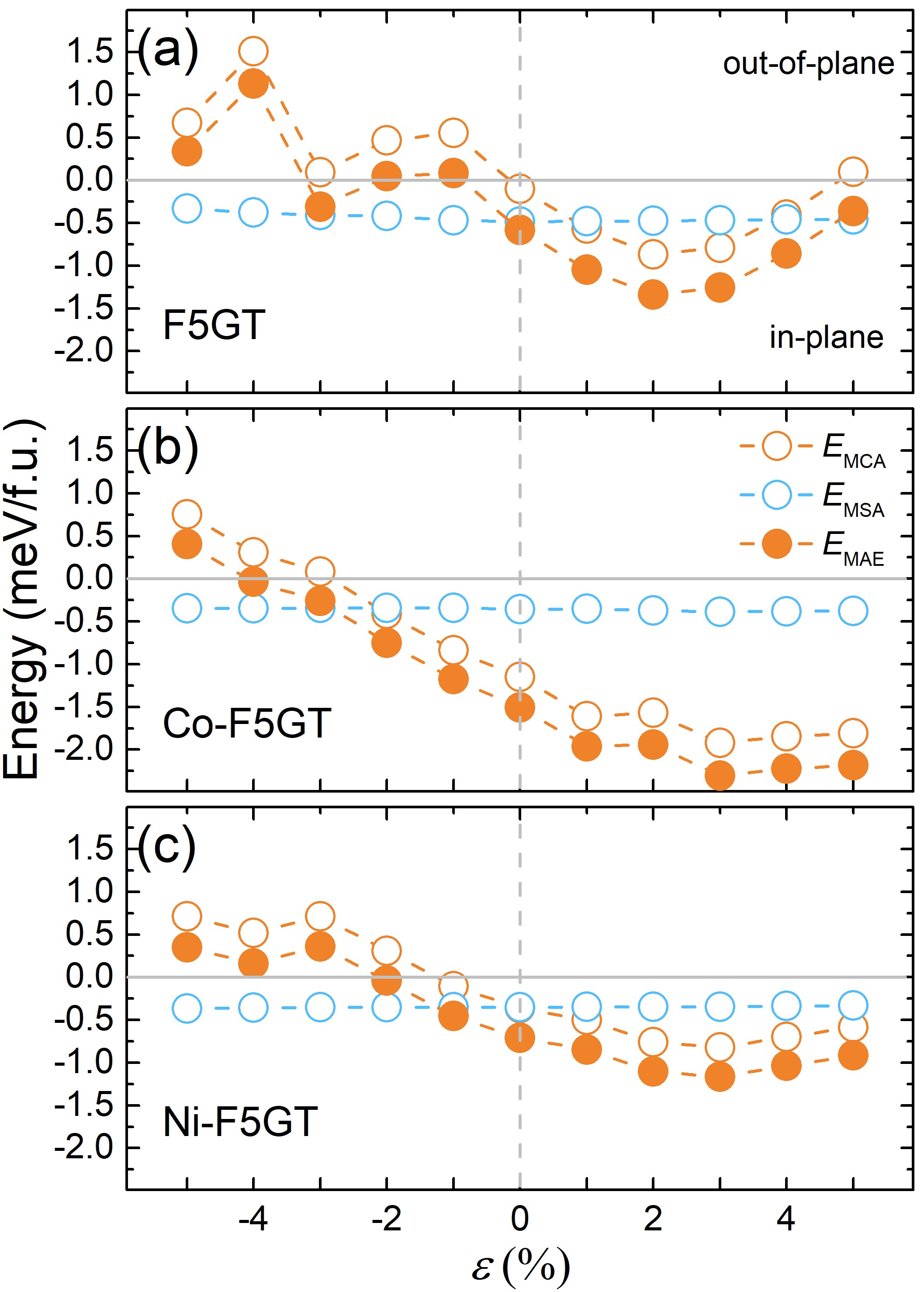}
	\caption{Strain-dependent $E_{\rm MCA }$, $E_{\rm MSA }$, and $E_{\rm MAE }$ of (a) F5GT, (b) Co-F5GT, and (c) Ni-F5GT monolayers.}
	\label{fig3}
\end{figure}

Fig. \ref{fig3} presents the effects of strain, ranging from $-5\%$ to $+5\% $, on the MCA, MSA, and MAE of the F5GT, Ni-F5GT, and Co-F5GT monolayers. Unsurprisingly, $E_{\rm MSA }$ is relatively insensitive to strain, showing minimal variation, leading the strain-dependent behavior of $E_{\rm MAE }$ to closely resemble that of $E_{\rm MCA }$. For simplicity, we focus on the behavior of $E_{\rm MAE }$ with varying strain. Under tensile strain, the magnitude of the in-plane MAE increases, while compressive strain leads to a decrease in $E_{\rm MAE}$, which can even reverse its sign, indicating a transition from in-plane to out-of-plane magnetization. Specifically, for the Co-F5GT monolayer, $E_{\rm MAE }$ varies from 0.40 meV/f.u. to -2.18 meV/f.u., with a notable peak of -2.31 meV/f.u. at $\varepsilon = 3\%$~(Fig. \ref{fig3}(b)). Similarly, the Ni-F5GT monolayer shows $E_{\rm MAE }$ varying from 0.35 meV/f.u. to -0.92 meV/f.u. (Fig. \ref{fig3}(c)). In contrast, the $E_{\rm MAE}$ of F5GT monolayer shows more osccilation, with $E_{\rm MAE }$ varying between 0.34 meV/f.u. and -0.36 meV/f.u.~(Fig. \ref{fig3}(a)). Remarkably, a minor compressive strain of $1\%$ is sufficient to switch the easy axis of magnetization in F5GT. Furthermore, with $\varepsilon = -4\%$, F5GT exhibits a relatively large out-of-plane magnetic anisotropy with a value of 1.13 meV/f.u.

\begin{figure}[t]
	\includegraphics[scale=0.35]{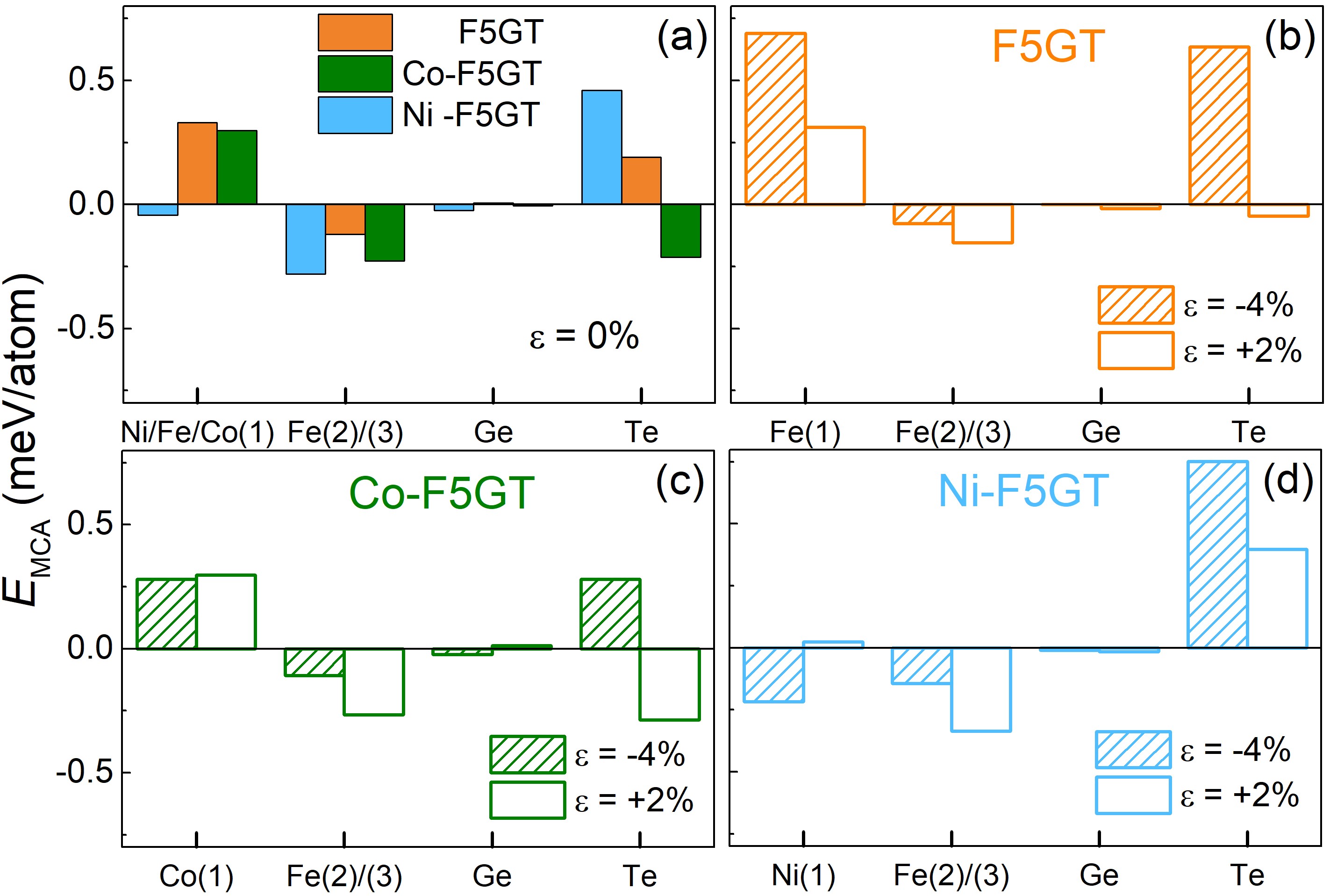}
	\centering
	\caption{(a) The atom-decomposed $E_{\rm MCA}$ for unstrained F5GT (orange solid bar graph), Co-F5GT (green solid bar graph), and Ni-F5GT (blue solid bar graph) monolayers. Panels (b), (c), and (d) show the atom-decomposed $E_{\rm MCA}$ of F5GT, Co-F5GT, and Ni-F5GT monolayers, respectively, under strain conditions of $\varepsilon = -4\%$ (shaded bar chart) and $+2\%$ (hollow-framed bar graph). Here Fe(1) denotes contributions from Fe(1)$_{up}$ and Fe(1)$_{dn}$, Fe(2)/(3) represents contributions from Fe(2) and Fe(3).}
	\label{fig4}
\end{figure}
\subsection{Origins of the tunable MAE}
In this section, we provide a detailed analysis of the strain-dependent MAE (only $E_{\rm MCA}$) through atomic decomposition and SOC matrix elements. The atomic contributions to $E_{\rm MCA}$ are estimated using the relation $E_{\rm MCA}$ $\approx$ $\frac{1}{2} \Delta E_{\rm SOC}$~\cite{2014Vladimir}, where $E_{\rm SOC}^{i}$ denotes the difference in SOC energies at atomic site $i$ between in-plane and out-of-plane magnetization. According to Eq. ~\ref{eq:eq1}, the $E_{\rm MCA }$ can be decomposed into contributions from individual atoms by considering SOC matrix elements formed by local orbitals of each atom. Fig. \ref{fig4}(a) illustrates the atomic decomposition of $E_{\rm MCA }$ for unstrained F5GT, Ni-F5GT, and Co-F5GT monolayers. For F5GT, the contributions to $E_{\rm MCA}$ from Fe(1) and Te atoms are negative, while Fe(2) and Fe(3) contribute positively. The competition between negative and positive contributions results in a relatively weak negative MCA for F5GT. In contrast, for Co-F5GT, the contribution from Co(1) remains similar to that of Fe(1) in F5GT, but the negative contributions from Fe(2) and Fe(3) increase, further enhanced by the contributions from Te atoms. Consequently, Co-F5GT exhibits a stronger in-plane MCA compared to F5GT. For Ni-F5GT, the contribution of Ni(1) to $E_{\rm MCA}$ is minimal. Conversely, Fe(2) and Fe(3) contribute significantly with a negative sign, while Te atoms contribute positively. This pattern of contributions results in a relatively weak in-plane magnetic easy axis for Ni-F5GT. Notably, the contribution from Ge atom is negligible in all cases.
 
\begin{figure*}[t]
	\includegraphics[scale=0.50]{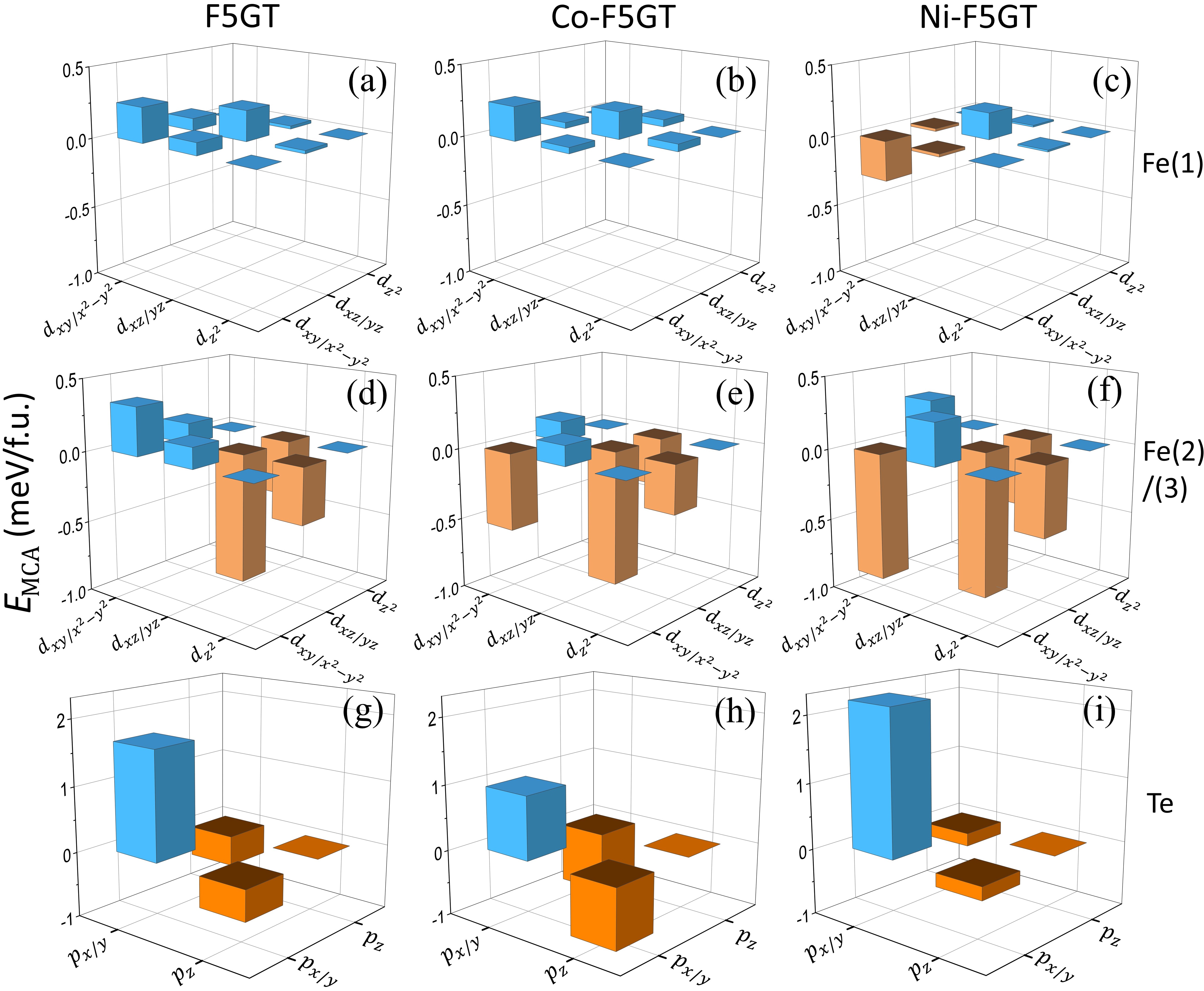}
	\caption{Orbital-resolved SOC matrix for unstrained (a)(d)(g) F5GT, (b)(e)(h) Co-F5GT, and (c)(f)(i) Ni-F5GT monolayers. Blue and orange denote positive and negative $E_{\rm MCA}$, respectively. }
	\label{fig5}
\end{figure*}
The variation in $E_{\rm MCA }$ is pronounced under both compressive strain at $\varepsilon = -4\%$ and tensile strain at $\varepsilon = +2\%$. To illustrate the impact of strain on $E_{\rm MCA }$, we compare these two cases. As depicted in Fig. \ref{fig4}(b), the dominant positive contributions from Te and Fe(1) atoms, in contrast to the smaller negative contributions from Fe(2)/(3) atoms, lead to the large out-of-plane MAE observed in F5GT at $\varepsilon = -4\%$. Conversely, as shown in Fig. \ref{fig4}(c), the significant in-plane MAE observed in Co-F5GT at $\varepsilon = +2\%$ arises from the competition between the large negative contributions of Te and Fe(2)/(3) atoms and the smaller positive contributions of Co(1) atoms. For Ni-F5GT (See Fig. \ref{fig4}(d)), the contribution from Ni(1) atoms remains minor across various strain conditions. The contributions from Fe(2)/(3) and Te atoms are substantial but with opposite signs, resulting in consistently low values of $E_{\rm MCA }$ for Ni-F5GT. As shown in Fig \ref{fig4}(b)~-~\ref{fig4}(d), the contribution of Te atoms to MAE is significant and highly sensitive to strain variations. Nevertheless, the overall magnitude and direction of MAE are determined by the combined effects of Fe(1), Fe(2), Fe(3), and Te atoms, rather than being predominantly determined by the Te atoms, as observed in Fe$_3$GeTe$_2$~\cite{2024kim}.

\begin{figure}[t]
	\includegraphics[scale=0.37]{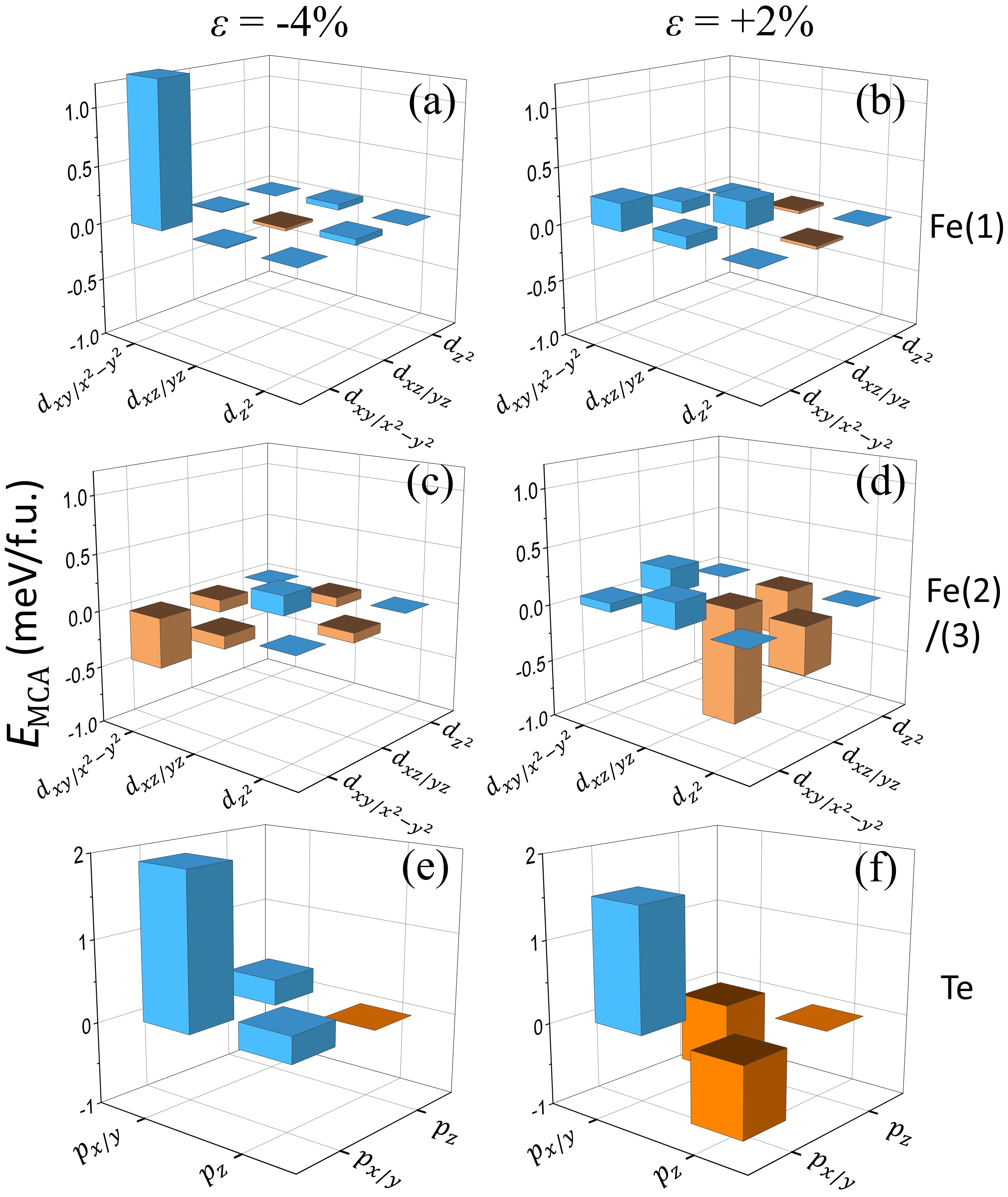}
	\centering
	\caption{Orbital-resolved SOC matrix for F5GT monolayer. Left and right panels display $\varepsilon = -4\%$ and $\varepsilon = +2\%$, respectively. Blue and orange denote positive and negative $E_{\rm MCA}$, respectively.}
	\label{fig6}
\end{figure}

To further elucidate the mechanisms underlying the tuning of the MAE in unstrained F5GT, Ni-F5GT, and Co-F5GT monolayers, we calculate and analyze the SOC matrix elements of Te, Fe, Ni(1), and Co(1) atoms, as illustrated in Fig. \ref{fig5}. In a system with trigonal symmetry, the {\it d} orbitals split into three distinct sets according to the magnetic quantum number {\it m}: $\left|{\it m}\right|=0$~({\it d}$_{z^2}$), $\left|{\it m}\right|=1$~({\it d}$_{xz/yz}$), and $\left|{\it m}\right|=2$~({\it d}$_{xy/x^2+y^2}$). The {\it p} orbital represented by $\left|{\it m}\right|=0$~({\it p}$_{z}$) and $\left|{\it m}\right|=1$~({\it p}$_{x/y}$).

For F5GT, the contribution from Fe(1) predominantly comes from  $ \left\langle {\it d}_{xz} \left\vert L_z\right\vert {\it d}_{yz} \right\rangle  $ and $ \left\langle {\it d}_{xy} \left\vert L_z\right\vert {\it d}_{x^2+y^2} \right\rangle  $~(See Fig. \ref{fig5}(a)). In contrast, for Fe(2)/(3) atoms, the contribution from  $ \left\langle {\it d}_{xz} \left\vert L_z\right\vert {\it d}_{yz} \right\rangle  $ and $ \left\langle {\it d}_{z^2} \left\vert L_x\right\vert {\it d}_{yz} \right\rangle  $ dominate, leading to a negative contribution to the $E_{\rm MCA}$ (Fig. \ref{fig5}(d)). For Te, the positive contribution primarily comes from $ \left\langle {\it p}_{x} \left\vert L_z\right\vert {\it p}_{y} \right\rangle  $, with a smaller negative contribution from $ \left\langle {\it p}_{z} \left\vert L_x\right\vert {\it p}_{y} \right\rangle  $ (Fig. \ref{fig5}(g)). 

For Co-F5GT, the contributions from Co(1) are similar to those from Fe(1) in F5GT~(Fig. \ref{fig5}(b)). However, the sign of the $E_{\rm MCA}$ arises from $ \left\langle {\it d}_{xy} \left\vert L_z\right\vert {\it d}_{x^2+y^2} \right\rangle  $ in Fe(2)/(3) is reversed (Fig. \ref{fig5}(e)). Additionally, the positive contribution from $ \left\langle {\it p}_{x} \left\vert L_z\right\vert {\it p}_{y} \right\rangle  $ in Te decreases~(Fig. \ref{fig5}(h)). Consequently, the pronounced in-plane MAE in Co-F5GT compared to F5GT can be attributed to reduced positive contribution from Te and increased negative contribution from Fe(2)/(3). 

For Ni-F5GT, the contributions from $ \left\langle {\it d}_{xy} \left\vert L_z\right\vert {\it d}_{x^2+y^2} \right\rangle  $ and $ \left\langle {\it d}_{xz/yz} \left\vert L_x\right\vert {\it d}_{xy/x^2+y^2} \right\rangle  $ negatively impact the $E_{\rm MCA}$ in Ni(1)~(Fig. \ref{fig5}(c)). Athough the $E_{\rm MCA}$ contributions from $ \left\langle {\it d}_{xy} \left\vert L_z\right\vert {\it d}_{x^2+y^2} \right\rangle  $ in Fe(2)/(3) also undergoes a sign change compared to F5GT~(Fig. \ref{fig5}(f)), the positive contribution from $ \left\langle {\it p}_{x} \left\vert L_z\right\vert {\it p}_{y} \right\rangle  $ in Te increases. As a result, the negative $E_{\rm MCA} $ in Ni-F5GT shows a slight increase compared to F5GT. The matrix elements for {\it d} orbitals of Te and the {\it p} orbitals of Fe are not included, as their contributions are negligible. 

Fig. \ref{fig6} presents the orbital-projected SOC matrices for Fe {\it d} and Te {\it p} orbitals. The left column depicts the matrices under compressive strain at $\varepsilon = -4\%$, while the right column displays those under tensile strain at $\varepsilon = +2\%$. As shown in Fig. \ref{fig6}(a), the contribution to the $E_{\rm MCA}$ from $ \left\langle {\it d}_{xy} \left\vert L_z\right\vert {\it d}_{x^2+y^2} \right\rangle  $ in Fe(1) is significantly more positive compared to the unstrained F5GT. For Fe(2)/(3), the contribution from $ \left\langle {\it d}_{xz} \left\vert L_z\right\vert {\it d}_{yz} \right\rangle  $ is reversed, resulting in a positive $E_{\rm MCA}$. However, the contributions from $ \left\langle {\it d}_{xy} \left\vert L_z\right\vert {\it d}_{x^2+y^2} \right\rangle  $ and $ \left\langle {\it d}_{xz/yz} \left\vert L_x\right\vert {\it d}_{xy/x^2+y^2} \right\rangle  $ change from positive to negative, as depicted in Fig. \ref{fig6}(c). Consequently, the negative $E_{\rm MCA}$ contribution from Fe(2)/(3) slightly decreases. For Te atoms, the positive contribution from $ \left\langle {\it p}_{x} \left\vert L_z\right\vert {\it p}_{y} \right\rangle  $ is notably enhanced, accompanied by a weaker positive contribution from $ \left\langle {\it p}_{z} \left\vert L_x\right\vert {\it p}_{y} \right\rangle  $, as depicted in Fig. \ref{fig6}(e). Therefore, under compressive strain ($\varepsilon = -4\%$), F5GT exhibits a relatively large positive $E_{\rm MCA}$. 

On the other hand, under tensile strain at $\varepsilon = +2\%$, the contributions from Fe(1) exhibit minimal variation~(Fig. \ref{fig6}(b)), while the decomposition of $E_{\rm MCA}$ from Fe(2)/(3) and Te are more straightforward. There is no sign change in the contributions, but the positive contribution from $ \left\langle {\it d}_{xy} \left\vert L_z\right\vert {\it d}_{x^2+y^2} \right\rangle  $ in Fe(2)/(3) decreases~(Fig. \ref{fig6}(d)). In the mean time, the negative contribution from $ \left\langle {\it p}_{z} \left\vert L_x\right\vert {\it p}_{y} \right\rangle  $ in Te increases (Fig. \ref{fig6}(f)). Therefore, at $\varepsilon = +2\%$, F5GT shows a slight increase in negative $E_{\rm MCA}$.
\begin{figure}[t]
	\includegraphics[scale=0.34]{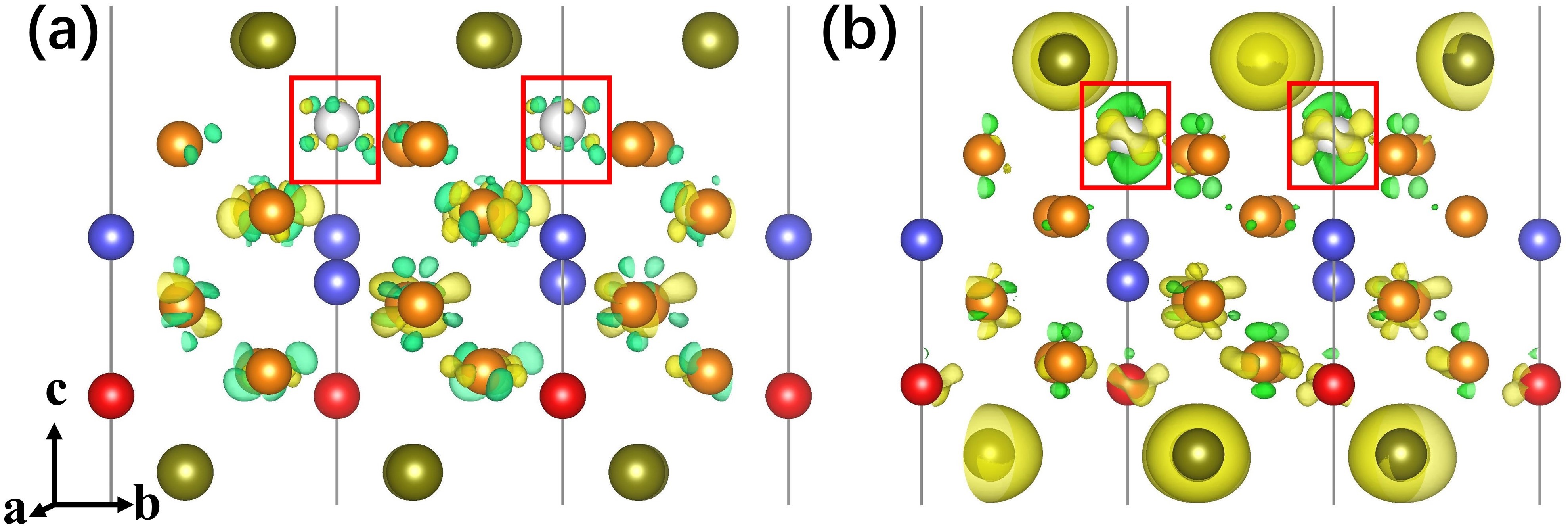}
	\centering
	\caption{(a) Differential partial charge density of the F5GT monolayer considering only the SOC of Fe, compared to no SOC. (b) Differential partial charge density of the F5GT monolayer considering the SOC of both Fe and Te atoms, compared to no SOC. The isosurface level in (a) is set to half of that in (b). The positions of Fe(1)$_{up}$ are labeled by red rectangles.}
	\label{fig7} 
\end{figure}
\begin{figure}[t]
	\includegraphics[scale=0.35]{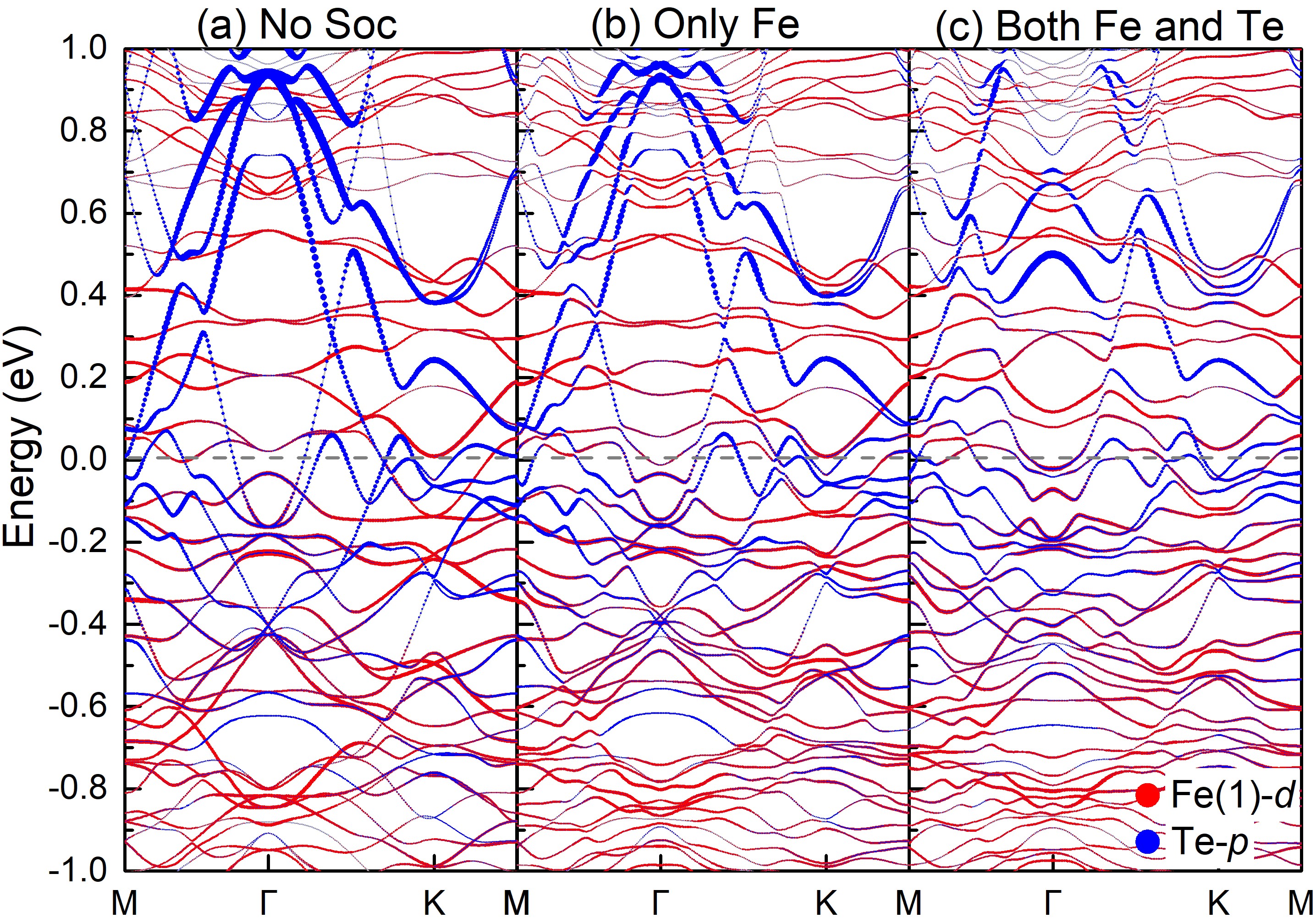}
	\centering
	\caption{The projected band structure of F5GT monolayer considering (a) no SOC, (b) only the SOC of Fe, and (c) the SOC of both Fe and Te atoms, respectively. The scaling factor for Te is set to be twice that of Fe(1) for clarity.}
	\label{fig8} 
\end{figure}

It is well-known that the MCA primarily originates from the coupling between the local spin and orbital moments of magnetic Fe atoms. However, Figs. \ref{fig4}--\ref{fig6} suggest that the SOC of non-magnetic Te atoms also plays a significant role. To investigate the physical origin of Te's effect on MCA, we further calculate the differential partial charge densities, defined as $\Delta_{\rm Fe}=n_{Fe} - n_0$ and $\Delta_{\rm FeTe}= n_{FeTe} - n_0$, respectively. Here, $n_0$, $n_{Fe}$, and $n_{FeTe}$ are charge densities of F5GT monolayer when neglecting all SOC, considering only SOC of Fe, and including SOC of both Fe and Te, respectively. As shown in Figs. \Ref{fig7}(a) and \ref{fig7}(b), the results demonstrate that incorporating the SOC of Te significantly affect charge density distribution, particularly at the Fe(1) sites, which can be attributed to the stronger hybridization between Fe(1) and Te atoms. Additionally, we calculate the electronic band structure of F5GT monolayer, as shown in Fig. \ref{fig8}. Compared to the band structures without SOC, when only SOC of the Fe atoms is considered, some of the crossing bands (Fe(1)-{\it d} and Te-{\it p} bands) exhibit slight separation, with minor changes in the shape and position of these bands~(see Fig. \ref{fig8}(a) and \ref{fig8}(a)). In contrast, when the SOC of both Fe and Te atoms is included, significant alterations in the bands near the $\Gamma$ point are observed, which can affect the distribution of electronic states of both Te and Fe (see Fig. \ref{fig8}(c)). Overall, the strong SOC of the heavier Te atoms induces significant alterations in their atomic states, which in turn affect the adjacent Fe {\it d} states. This interplay likely explains the non-negligible contribution of Te to the MCA.

\subsection{Discussion}
We briefly discuss potential approaches for applying in-plane strain to the F5GT monolayer. Due to the challenges of directly introducing strain in freestanding 2D materials, a common approach is to transfer them onto a substrate, where strain can be applied by controlling substrate deformation. Flexible polymer substrates, such as polyethylene terephthalate (PET), polyvinyl alcohol (PVA), and polydimethylsiloxane (PDMS), are particularly effective for inducing mechanical strain due to their high elasticity and resilience. For example, by bulging a flexible PDMS substrate upward, strains in the range of 2.2--2.6$\%$ (measured by photoluminescence) and 2.9--3.5$\%$ (measured by Raman spectroscopy) can be applied to an attached MoS$_2$ monolayer~\cite{2017nanolettersyang}. Similarly, Weerahennedige et al. demonstrate that strains ranging from -2.1$\%$ to 2.5$\%$ can be introduced in Fe$_3$GeTe$_2$ flakes on PET substrate using a bending device~\cite{2024sai}. Additionally, encapsulating monolayer WSe$_2$ in a PVA substrate has enabled tensile strain as high as 5.86$\%$, attributed to the strong adhesion and interaction between PVA and WSe$_2$~\cite{2020ncli}. Piezoelectric substrates, such as lead zinc niobate-lead titanate (PZN-PT), are well-suited for fine-tuned, real-time modulation due to their electrostrictive properties. For instance,  an electric field of 120 kV/cm applied across PZN-PT can induce a strain of approximately 1.7$\%$, owing to its large piezoelectric coefficients~\cite{1997jappark}. Strain engineering can also be achieved through lattice mismatch during the epitaxial growth of 2D materials. In the MoS$_2$-WSe$_2$ lateral heterostructure, for example, maximum tensile strain of $1.59\pm0.25\%$ and compressive strain of $1.1\pm0.18\%$ can arise in the MoS$_2$ region due to lattice mismatch~\cite{2015scienceli}.
Overall, various substrate types, including flexible polymers, piezoelectric materials, and 2D heterostructures, provide distinct advantages for inducing strain in F5GT monolayers, each supporting specific strain ranges.

\section{Conclusions}

In summary, the strain-induced tunability of magnetic anisotropy in monolayers of F5GT, Co-F5GT, and Ni-F5GT has been systematically investigated based on first-principles calculations. Our calculations reveal that the total magnetic moment of these materials decreases under compressive strain while increases under tensile strain. The F5GT monolayer exhibits a weak in-plane magnetic anisotropy, which can be significantly enhanced by Co doping, resulting in a roughly threefold increase from -0.58 meV/f.u. to -1.51 meV/f.u. Moreover, inducing strain proves to be an effective method to tune the MAE. A sign change in MAE occurs at $ \varepsilon = -1\%$, indicating a transformation from in-plane to out-of-plane magnetization, while $4\%$ compressive strain further enhances the out-of-plane MAE, with a value of 1.13 meV/f.u. SOC matrix analysis reveals that the pronounced in-plane MAE in Co-F5GT is attributed to a reduction in the positive contribution from  $ \left\langle {\it p}_{x} \left\vert L_z\right\vert {\it p}_{y} \right\rangle  $  of Te and an increase in the negative contribution from  $ \left\langle {\it d}_{xy} \left\vert L_z\right\vert {\it d}_{x^2+y^2} \right\rangle  $  of Fe(2)/(3). The effect of compressive strain is primarily due to the significant increase in the positive contribution from  $ \left\langle {\it d}_{xy} \left\vert L_z\right\vert {\it d}_{x^2+y^2} \right\rangle  $ of Fe(1). These findings underscore the tunability of magnetic properties in F5GT, providing valuable insights for the design of future spintronic applications.

\section*{Acknowledgements}
Xunwu Hu is grateful to Yusheng Hou for his useful discussions. Work at Sun Yat-Sen University was supported by the National Key Research and Development Program of China (Grants No. 2022YFA1402802, 2018YFA0306001), and the Guangdong Basic and Applied Basic Research Foundation (Grants No. 2022A1515011618), and the National Natural Science Foundation of China (Grants No. 92165204, No. 11974432), Shenzhen International Quantum Academy (Grant No. SIQA202102), Leading Talent Program of Guangdong Special Projects (201626003), Fundamental Research Funds for the Central Universities, Sun Yat-sen University (Grants No. 23ptpy158), Guangdong Provincial Key Laboratory of Magnetoelectric Physics and Devices (No. 2022B1212010008).

%

\end{document}